\documentclass[conference]{IEEEtran}
\IEEEoverridecommandlockouts
\usepackage{cite}
\usepackage{amsmath,amssymb,amsfonts}
\usepackage{algorithmic}
\usepackage{graphicx}
\usepackage{textcomp}
\usepackage{xcolor}
\usepackage{balance}

\DeclareRobustCommand*{\IEEEauthorrefmark}[1]{%
  \raisebox{0pt}[0pt][0pt]{\textsuperscript{\footnotesize #1}}%
}

\def\BibTeX{{\rm B\kern-.05em{\sc i\kern-.025em b}\kern-.08em
    T\kern-.1667em\lower.7ex\hbox{E}\kern-.125emX}}
\begin{document}

\title{The Effect of Explicit Structure Encoding of Deep Neural Networks for Symbolic Music Generation
}
\author{\IEEEauthorblockN{Ke Chen\IEEEauthorrefmark{1}\,\IEEEauthorrefmark{4},
Weilin Zhang\IEEEauthorrefmark{2}\,\IEEEauthorrefmark{4}, Shlomo Dubnov\IEEEauthorrefmark{3},
Gus Xia\IEEEauthorrefmark{4} and Wei Li\IEEEauthorrefmark{1} \\
\IEEEauthorrefmark{1}School of Computer Science, Fudan University, China\\
\IEEEauthorrefmark{2}Department of Computer Science, University of Illinois Urbana-Champaign, USA\\
\IEEEauthorrefmark{3}Department of Music, University of California San Diego, USA\\
\IEEEauthorrefmark{4}Department of Computer Science, NYU Shanghai, China\\
kchen15@fudan.edu.cn,
weilinz2@illinois.edu,
sdubnov@ucsd.edu,
gxia@nyu.edu,
weili-fudan@fudan.edu.cn}}

\maketitle

\begin{abstract}
With recent breakthroughs in artificial neural networks, \textit{deep generative models} have become one of the leading techniques for computational creativity. Despite very promising progress on image and short sequence generation, symbolic music generation remains a challenging problem since the structure of compositions are usually complicated. In this study, we attempt to solve the melody generation problem constrained by the given chord progression. In particular, we explore the effect of explicit architectural encoding of musical structure via comparing two sequential generative models: LSTM (a type of RNN) and WaveNet (dilated temporal-CNN). As far as we know, this is the first study of applying WaveNet to symbolic music generation, as well as the first systematic comparison between temporal-CNN and RNN for music generation. We conduct a survey for evaluation in our generations and implemented \textit{Variable Markov Oracle} in music pattern discovery. Experimental results show that to encode structure more explicitly using a stack of dilated convolution layers improved the performance significantly, and a global encoding of underlying chord progression into the generation procedure gains even more. \\
\end{abstract}


\section{Introduction}
\textit{Automated music generation} has always been one of the principal targets of applying AI to music.  With recent breakthroughs in artificial neural networks, \textit{deep generative models} have become one of the leading techniques for automated music generation \cite{Briot}, and many systems have generated more convincing results than traditional rule-based methods \cite{Loy}. For the examples of re-generating J.S. Bach's work alone, we have seen \cite{BachBot} - \cite{CNNBach}. 

Despite these promising progress, people still struggle to generate well-structured music. It is worth noting that most successful cases of automatic music compositions were limited to Bach, and at least for non-experts the structure of Bach's compositions is rather local and easy to perceive compared to many other composers. In other words, automatic composition remains a challenging problem since music structures, for most compositions, are complicated and involve long-term dependencies. To solve this problem, some studies imposed structural restrictions \cite{Lattner} - \cite{StructureNet} on the final output. However, such post-processing restrictions usually conflict with the generating procedure and require tedious parameter tuning in order to make the algorithm converge. It makes more sense to embed the notion of music structure into the model architecture and generative procedure. 

In this study, we chose the task in generating melody constrained by given chord progression. As discussed and practiced by \cite{Cope}, the generation of music by computers is considered as music computational creativity. Solving this problem will show the importance of model choices and data representations in \textit{deep generative model}. We did a systematic comparison between two main-stream approaches of handling music structure representation using two sequential representation generative models: LSTM (a type of RNN) and WaveNet (dilated temporal-CNN). The former encodes structure purely implicitly by the memory of hidden states, while the latter adds more explicit structured dependency via a larger receptive field of dilated convolutions. In terms of the dependencies between hidden variables, the relationship between LSTM and WaveNet is analogous to the one between a first-order autoregressive moving average (ARMA) model and a higher-order moving average (MA) model. From a signal processing perspective, the output signals of LSTM and ARMA models depend on both history input and output signals, while the output signals of WaveNet and MA models solely depend on history inputs. To our knowledge, this is the first systematic comparison between temporal-CNN and RNN for symbolic music application. 

We focus on symbolic music generation because music structure information is richer at the composition level than the performance and acoustic level \cite{position}. As far as we know, this is the first attempt in applying WaveNet to symbolic music generation (The name of WaveNet implies its usage on audio applications, but in theory the temporal-CNN architecture can also be used for symbolic generation). Similarly to other studies \cite{DeepBach} \cite{Hild}, we use chord progression as the global input for both models and turn the task into modeling the conditional distribution of music composition given chords. We present a novel way of encoding chords and melody in a staggered representation. This effectively combines aspect of different time scales of chords and melody in music, learns simultaneously temporal delayed dependencies between melody over past and next two bars, and also learns harmonic-melodic simultaneous relations within every two bars of music. Such manipulation makes sense on a real composition scenario since in this context musicians rarely do purely-free improvisation (unconditioned generation) and almost always rely on a pre-defined guide (e.g. figured bass, chord progression, lead sheet, etc.) which encodes high-level music structure information.

In order to evaluate the performance of the neural model, we conducted a subjective survey to evaluate the quality of generated music. Human judgment takes into account, unconsciously, not only the local musical statistics, but also builds anticipations that keep track of long music structure, such as recognition of salient motifs and their patterns \cite{AMMS}. To date, most of the evaluation metrics for neural music models were done in terms of immediate prediction error, incapable of capturing longer terms salience structures. In order to be able to see how well the neurally generated music is able to learn such structure, we applied an \textit{Information Dynamics} analysis developed by \cite{VMO} for music pattern discovery. We applied this analysis to several musical music versus model-generated examples. Experimental results show that in terms of \textit{Information Dynamics} ability for encoding of longer terms music structure, using dilated convolution layers improved the performance significantly. Moreover, we found that the results further improve when we incorporated the complete chord progression into the generation procedure rather than merely considering partial past chords. Our results show that repetition patterns will be found more clearly in the generation if we incorporate the global structure into our inputs.

In the next section, we present related works. We describe the methodology in Section III and show the experimental results in Section IV. We discuss several important discoveries in Section V and finally come to the conclusion in Section VI.

\section{Related Work}

\subsection{WaveNet for Sound Generation}
WaveNet \cite{Wavenet} was first introduced by Google Deepmind as a generative model for raw audio. Since then, we have seen many follow-up studies. Most works focus on two aspects: improving the speed of WaveNet, and applying WaveNet to audio-related applications. Parallel WaveNet \cite{parallelwav} speeds up the generation process, and Fast WaveNet \cite{fastwav} reduces the time complexity. WaveNet was used in many aspects of raw audio generation as auto-encoder and audio synthesizer. Applications include timbre style generator \cite{wavtimbre},  voice conversion \cite{Kobayashi}, speech synthesis \cite{Tamamori} \cite{Shen}, speech enhancement \cite{Qian}, cello performance synthesizer \cite{manzelli2018end}, and speech denoising \cite{denoise}. Most convincing results were achieved via adding conditions as an extra input. For example, the neural audio synthesizer by WaveNet auto-encoders \cite{wavtimbre} add pitch conditioning during training. 

\subsection{LSTM for Music Generation}
Many music generation works by deep neural networks start with unconditional (monophonic) symbolic melody generation. The initial work \cite{Todd} implemented the Back-Propagation Through Time (BPTT) algorithm and used melody and duration representation as the training input for generation.  Since generation from single melody can be unstructured, follow up works usually includes conditions on chords or other musical features to guide the generation process. 

With Recurrent Neural Network (RNN) \cite{rnn} and its advanced versions (LSTM and GRU) \cite{lstm} \cite{gru} came out, long-term dependency can be captured for music generation. The work by \cite{lstmblue} demonstrated that RNNs is capable of revealing some higher-level information in melody generation. They tested the Blues improvisation performance of LSTM by inputting note slices in real time. The work by \cite{coca} defined several measurements (Tone division, Mode, Number of Octaves, etc.) and create melody sequences by RNN by varies inspirations. The unit selection method \cite{Bretan} took a series of measures in music as a unit and used a deep structured semantic model (DSSM) with LSTM to predict future units, instead of directly generate essential elements like notes.  

An important variation is the bidirectional architecture. DeepBach \cite{DeepBach} introduced an innovated bidirectional RNN for music harmonization. However, the main purpose of DeepBach is harmonization, not to use bidirectional neural networks for music generation. This work inspired us to use Bi-LSTM for conditioned melody generation. 

\section{Methodology}
\subsection{Problem Definition}
For music piece of length $T$, given the melody until time point $t$ ($t$$<$$T$) and the chords for the whole piece, we aim to generate the melody from $t$ to $T$ under the chord conditioning. In other words, we have two sequences for the input, the melody sequence and the conditioning sequence. During the generation process, the chord condition is given at each time step for a guiding purpose, and the final output shall both keep the melodic flow and interacts with chords. 

Such conditioned generation problem is shown in Fig. 1. The black notes $M_{1:t}$ and $C_{1:T}$ represent all the inputs. The blue notes $M_{t+1:T}$ represent the predicted sequence. The upper track is the melody track, and the lower track is the chord track.
The conditional probability distribution for one-way model is defined as:
\begin{equation}\label{equ1}
p\left( M_{t+1:T} \mid C_{1:T} \right) = \prod_{i=t+1}^{T} p\left(m_i \mid m_1, \dots ,m_{i-1} , c_1, \dots , c_{i} \right)
\end{equation}

where $m_i$ is the generated melody at time step $i$, $c_i$ is the condition at time step $i$. 

\begin{figure}[h]
  \centering
  \includegraphics[width=0.98\linewidth]{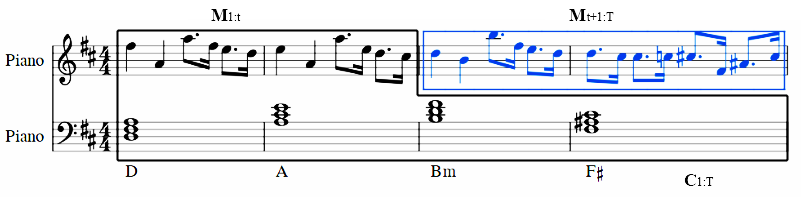}
  \caption{An illustration of staggered representation in our generation problem}
  \label{fig:wavenet_archi}
\end{figure}

\begin{figure}[ht]
  \centering
  \includegraphics[width=0.9\linewidth]{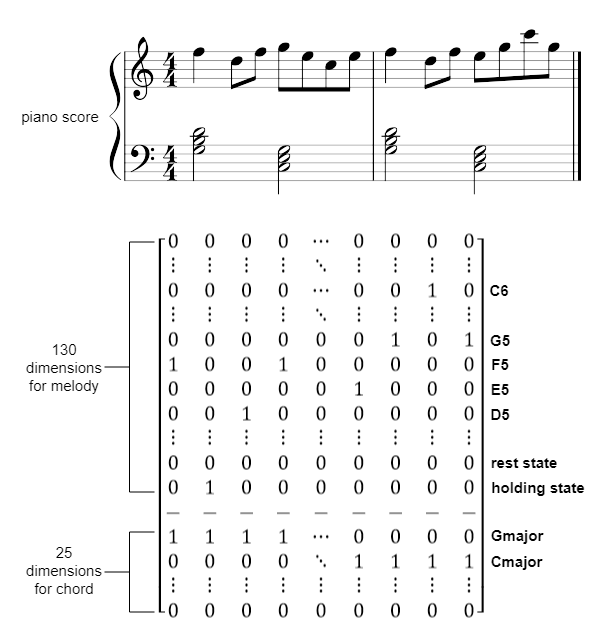}
  \caption{Data representation of LSTM models.}
  \label{fig:lstm_data}
\end{figure}

\subsection{Data Representation}

For LSTM, we represent the input as a vector $V$, which consists of two one-hot vectors $M$ and $C$, representing melody and chord respectively. 
\begin{equation}\label{equ1}
V = (M, C)
\end{equation}

As shown in Fig. 2, the bottom is the original symbolic file in music score. On top of Fig. 2, the two tracks are the representation of vector sequences with melody part and chord part. Note that the signs in the right side do not represent the actual order in the representation. We use a $T\times155$ dimension vector (130 dimensions for melody and 25 dimensions for chord) to represent a piece of the music with the number of time steps denoted as $T$. In the melody vector $M$, the 130 dimensions include pitch, duration, and the rest sign. We use holding state and rest state mentioned in \cite{musicvae}. The first 128 dimensions in $M$ represents pitch value from 0 to 127. Dimension 129 is the rest state, which implies that the note is empty. The last dimension is the holding state, which represents the duration of the previous pitch. 

Similar to the melody vector $M$, the first 24 dimensions in chord vector $C$ represent the most common 12 major chords and 12 minors chords regardless of the inversion (i.e. different root note in one chord). The last dimension is the none chord sign $NC$. For a chord that is not in the most common 24 chords, we match it to one of the most common chords that share the largest number of same pitches. For example, C-major7 (C7) matches C-major, C-minor7 (Cm7) matches C-minor and C-augment (Caug) matches C-major. The chord vector $C$ does not have the holding state since melody generation is more related to chord value than duration. 

\begin{figure}[ht]
  \centering
  \includegraphics[width=0.98\linewidth]{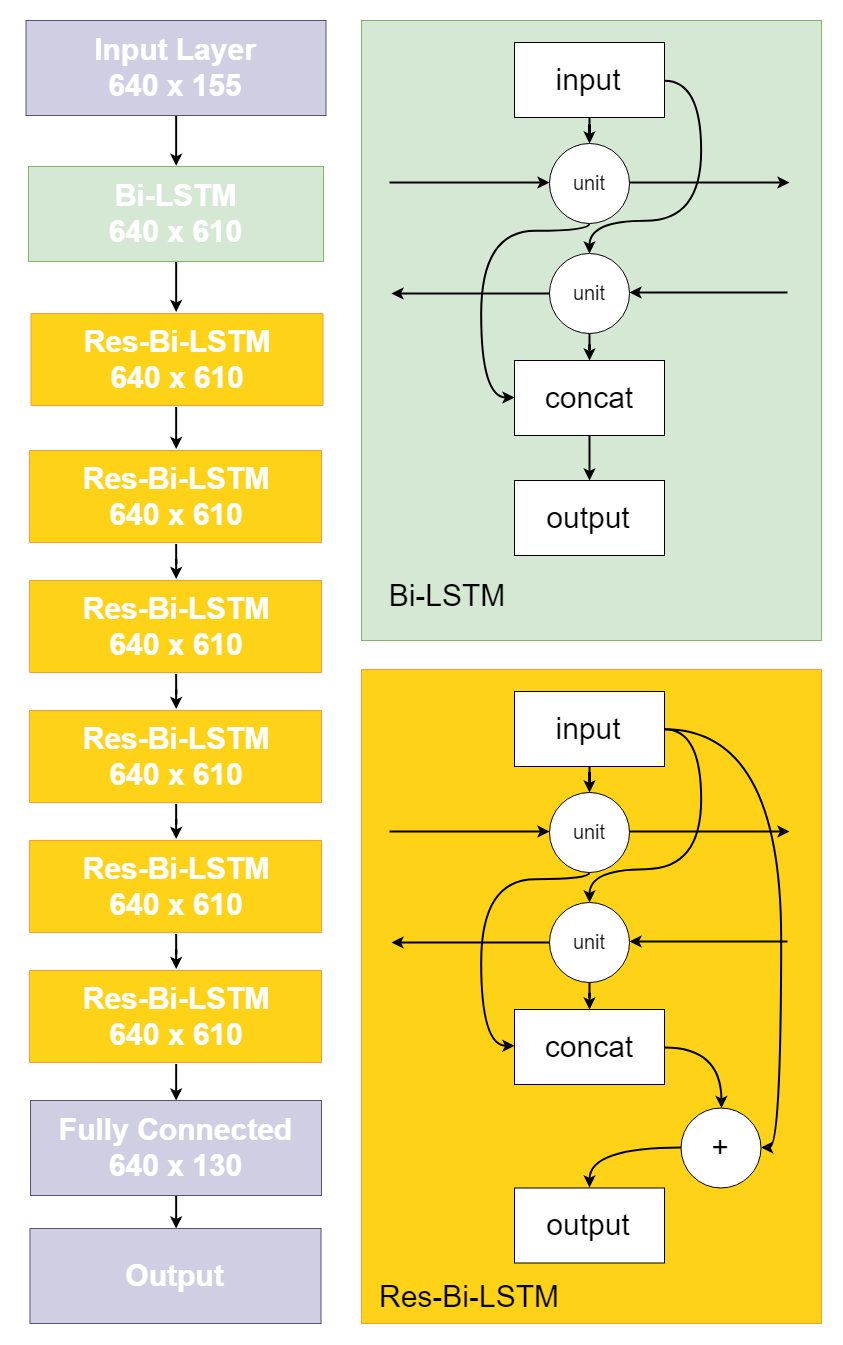}
  \caption{The bidirectional LSTM model.}
  \label{fig:bilstm}
\end{figure}
For the WaveNet model, the melody vector $M$ has dimension $T\times128$. Each channel represents a pitch, and the lowest pitch 0 means a rest. If the pitch value is the same in two or more consecutive time steps, we consider them as a sustaining note. For the condition vector, we hash all chords into 24 types, and the chord input is a one-hot vector over the hash table. Each input chord vector $C$ has the shape $T\times25$ (the additional channels means $NC$), where the number of time steps $T$ is the same as the corresponding melody vector. 

\subsection{LSTM Architecture}
In our LSTM model, we start with a one-way model and further develop it into a bidirectional one. As present above, we adapt our data to fit into both the unidirectional model and the bidirectional model. 

LSTM is consists of four gates: a cell gate $c$, an input gate $i$, an output gate $o$ and a forget gate $f$:
\begin{equation}\label{equ1}
{f_{t}} = \sigma_{g}(W_{f}x_{t}+U_{f}h_{t-1}+b_{f})
\end{equation}
\begin{equation}\label{equ1}
{i_{t}} = \sigma_{g}(W_{i}x_{t}+U_{i}h_{t-1}+b_{i}),
\end{equation}
\begin{equation}\label{equ1}
{o_{t}} = \sigma_{g}(W_{o}x_{t}+U_{o}h_{t-1}+b_{o}),
\end{equation}
\begin{equation}\label{equ1}
{c_{t}} = f_{t}\odot c_{t-1} + i_{t}\odot \sigma_{c}(W_{c}x_{t}+U_{c}h_{t-1}+b_{c}),
\end{equation}
\begin{equation}\label{equ1}
{h_{t}} = o_{t}\odot\sigma_{h}(c_{t})
\end{equation}

The detailed structure of our bidirectional LSTM model is illustrated in Fig. 3. The left is the input layer followed by 7-layer bidirectional LSTM layers and the fully connected layer. To speed up convergence, we referred to \cite{resnet} and use the skip-connection in all recurrent layers except the first one, which is shown on the right side.

One thing must be noticed is that the difference between the unidirectional and the bidirectional LSTM model lies in the amount of chord information. In the unidirectional model, we have only previous chord progression when generating the current note. However, the bidirectional model allows us to add the whole chord progression (i.e. the global structure) to the generation procedure. In that, the conditional probability for bidirectional LSTM model should be revised as:
\begin{equation}\label{equ1}
p\left( M_{t+1:T} \mid C_{1:T} \right) = \prod_{i=t+1}^{T} p\left(m_i \mid m_1, \dots ,m_{i-1} , c_1, \dots , c_{T} \right)
\end{equation}

We should note that the chord condition $c_1,...,c_{T}$ is different from the one-way formula $c_1,...,c_{i}$. The bidirectional model takes the complete chord progression as the condition. 

\subsection{WaveNet Architecture}
WaveNet proposed to use a stack of dilated temporal convolution layers \cite{dilated} for sequential prediction. The original paper also introduced the conditioning feature to guide music generation. For instance, the model can add personal information as a global condition to generate speech from certain people. 

We propose to apply WaveNet to symbolic music generation. The unconditioned model with the input melody vector $m$, and the activation function in dilation layer $k$ is:
\begin{equation}\label{equ?}
{z} = \mbox{tanh}(W_{f,k} \ast {m}) \odot \sigma(W_{g,k} \ast {m})
\end{equation}
where $\ast$ represents a dilated convolution operator, $W_{f,k}$ and $W_{g,k}$ are the learnable parameters in the convolution layer, and $\odot$ is a piecewise multiplication operator. 

We generate melody with conditioning on chords. We add the embedded chord vector as local conditioning in WaveNet. The activation function at layer $k$, with the embedded chord condition vector $c$ and the melody vector $m$: 
\begin{equation}\label{equ?}
{z} = \mbox{tanh}(W_{f,k} \ast {m} + V_{f,k} \ast {c}) \odot \sigma(W_{g,k} \ast {m} + V_{g,k} \ast {c})
\end{equation}
where the first $\ast$ in both parentheses represents a dilated convolution operator with $W_{f,k}$ and $W_{g,k}$ as the learnable parameters. The second $\ast$ in both parentheses represent a 1*1 convolutional layer, with $V_{f,k}$ and $V_{g,k}$ as the learnable parameters. We model the chord conditioning vector as the same length with the melody tensor. In this way, we assign chord to the melody sequence at each time step for monitoring purpose. Conditioning on chords guide the music generation process to include more musical structures, which improves the quality of generated music. The overall architecture and a stack of dilated convolutions is shown in Fig. 4 and Fig. 5. 

For all models, we use the cross-entropy loss between the generation melody and original sample melody as the loss function.

\begin{figure}[t]
  \centering
  \includegraphics[width=1.0\linewidth]{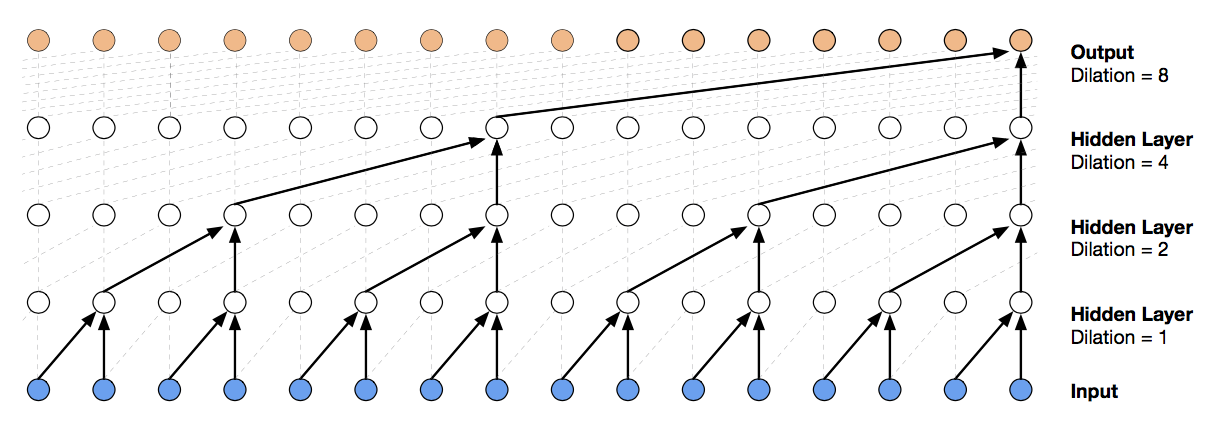}
  \caption{A stack of dilated temporal convolution layers, reproduced from \cite{Wavenet}.}
  \label{fig:dilatedconv}
\end{figure}

\begin{figure}[t]
  \centering
  \includegraphics[width=1.0\linewidth]{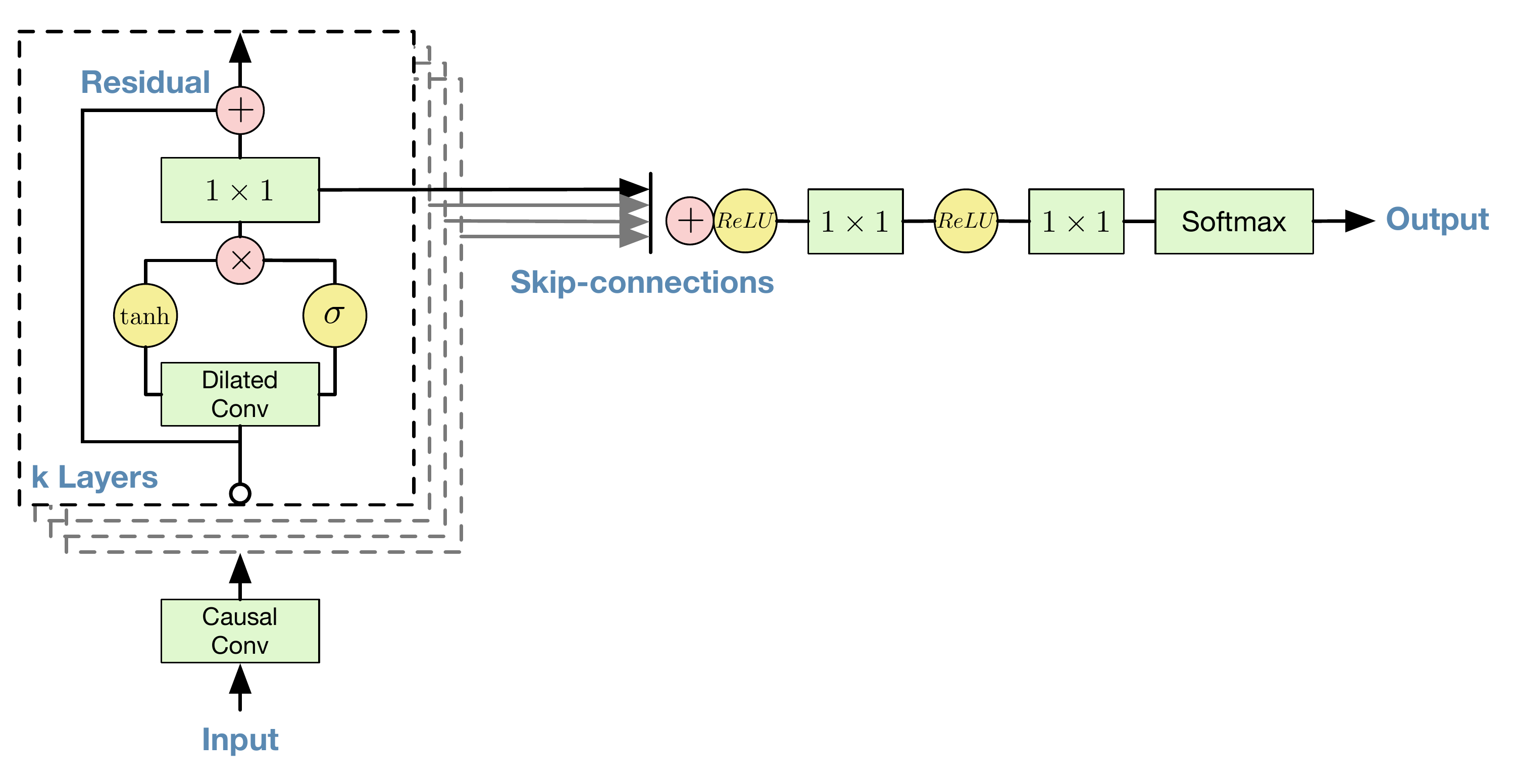}
  \caption{The WaveNet architecture, reproduced from \cite{Wavenet}.}
  \label{fig:wavenet_archi}
\end{figure}

\section{Experiments}
\subsection{Dataset}
We use the Nottingham Database \cite{Nottingham}, which consists of 941 folk songs, and each contains both melody and chords. We use 631 songs for training and use data augmentation, switching all the available songs to 12 major(or minor) tonalities, on the training set.  

We normalized all the music files to a fixed 120 bpm (beats per minute) to make better alignments in data. We set the frame size to be 1/16 beat(i.e. 0.5/16=0.03125 second) to sample most songs without any quantization error. In the case of triplets, our rule is to set them to be uneven durations. For example, a 5/16 + 6/16 + 5/16 group, we turn such groups back to ternary rhythmic patterns when we convert them to piano score.

\subsection{Survey}
We conducted a survey on audiences to compare the performance of the three proposed models (LSTM, Bi-LSTM, and WaveNet). During the survey, each audience listened to 3 groups of music pieces. Each group contains 3 samples generated by 3 models from a same piece of melody (3 x 3 = 9 samples in total). All samples contain 20 beats of original music followed by 20 beats of computer-generated music. After listening to each sample, audiences were asked to grade it in a continuous scale from 1 (low) to 5 (high). Namely, the final grade considers the following three criteria: \\
1. Interactivity: 
   Do the chords and melodies interact with each other well? \\ 
2. Complexity:
  Are the notes pattern complex enough to express the theme? \\
3. Structure:
  Can you notice some repetitions, forwards, variations in the sample?

\subsection{Hypothesis Test}
We performed the ANOVA \cite{anova} and two-sample t-test on all pairs of different models. 

The null hypothesis of ANOVA is that there is no difference in performance between the three models. Formally:

\begin{equation}\label{equ?}
H_{0} : \mu_{A} =  \mu_{B} = \mu_{C}
\end{equation}
the alternative hypothesis is that:
\begin{equation}\label{equ?}
H_{1}:  \exists i, j  \in \{A, B, C\}: \mu_{i} \neq  \mu_{j}
\end{equation}
The null hypothesis of the t-test is that there is no difference in performance between the two test models. Formally:
\begin{equation}\label{equ?}
H_{0} : \mu_{i} =  \mu_{j}
\end{equation}
the alternative hypothesis is that:
\begin{equation}\label{equ?}
H_{1}:  \mu_{i} \neq  \mu_{j}
\end{equation}

\subsection{Survey Evaluation}
A total of $n$ = 106 people (42 females and 64 males) have completed the survey. 69.81\% of them has experiences in music. The aggregated results are shown in Table 1 and Fig. 6. Below is the result table for the within-ANOVA test and the T-tests.  

We see that all p-values are smaller than 0.05. Therefore, the performance difference between the three models on conditioned melody generation is statistically significant . 

In Fig. 6, the height of the bars represent the means of the ratings, and the error bars represent the mean squared error (MSE).

WaveNet performs better than the unidirectional LSTM model. This result implies that the explicit dependency in dilated convolutions performs better than the implicit dependency in LSTM. The bidirectional LSTM is even better than WaveNet, and in our tests it is the best model. This result shows that embedding future chords in encoder largely improves the model performance. We will discuss the results further in Section V.

\subsection{Pattern Discovery by VMO}
We implemented \textit{Variable Markov Oracle} from \cite{VMO} to illustrate disparities of models in generating music patterns and repetition structures.

\begin {table}[h]
\setlength{\tabcolsep}{7mm}
\caption {Result Table} \label{tab:title} 
\begin{center} 
\begin{tabular}{ |c|c|}
 \hline
 ANOVA p-value & 3.02e-16  \\ 
  \hline
 T-test p-value 1, 2& 4.10e-16  \\ 
  \hline
 T-test p-value 1, 3& 0.017  \\ 
  \hline
 T-test p-value 2, 3& 5.29e-10  \\ 
 \hline 
\end{tabular}
\end{center} 
\end {table}

\begin{figure}[h]
  \centering
  \includegraphics[width=0.98\linewidth]{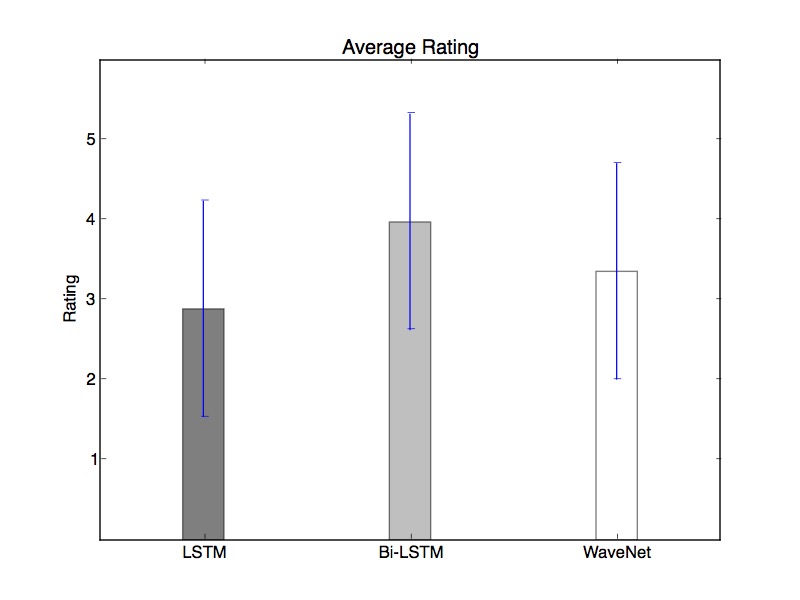}
  \caption{The subjective evaluation results of the ratings on three models. Different colors represent different models. }
  \label{fig:barchart}
\end{figure}
The \textit{Variable Markov Oracle} data structure can detect the repeated suffixes, or music patterns within a time series. Given a symbolic sequence $Q=q_{1},q_{2},...,q_{T}$, the VMO carries three kinds of links: forward link, suffix link(\textit{sfx}) and reverse suffix link(\textit{rsfx}). A suffix link of each time state $t$ is the starting point of the longest repeated suffix(\textit{lrs}) of the given Sequence $\{q_{1},q_{2},...,q_{t}\}$ . A reverse suffix link is the suffix link in a reverse direction.

As shown in Fig. 7 by \cite{VMO}, an example of the VMO structure in a symbolic signal sequence is provided. Solid arrows represent forward links and dashed arrows are suffix links. The visualization in the bottom part shows how the repetition parts are detected.
\\ \\ \\ \\

\begin{figure}[h]
  \centering
  \includegraphics[width=0.98\linewidth]{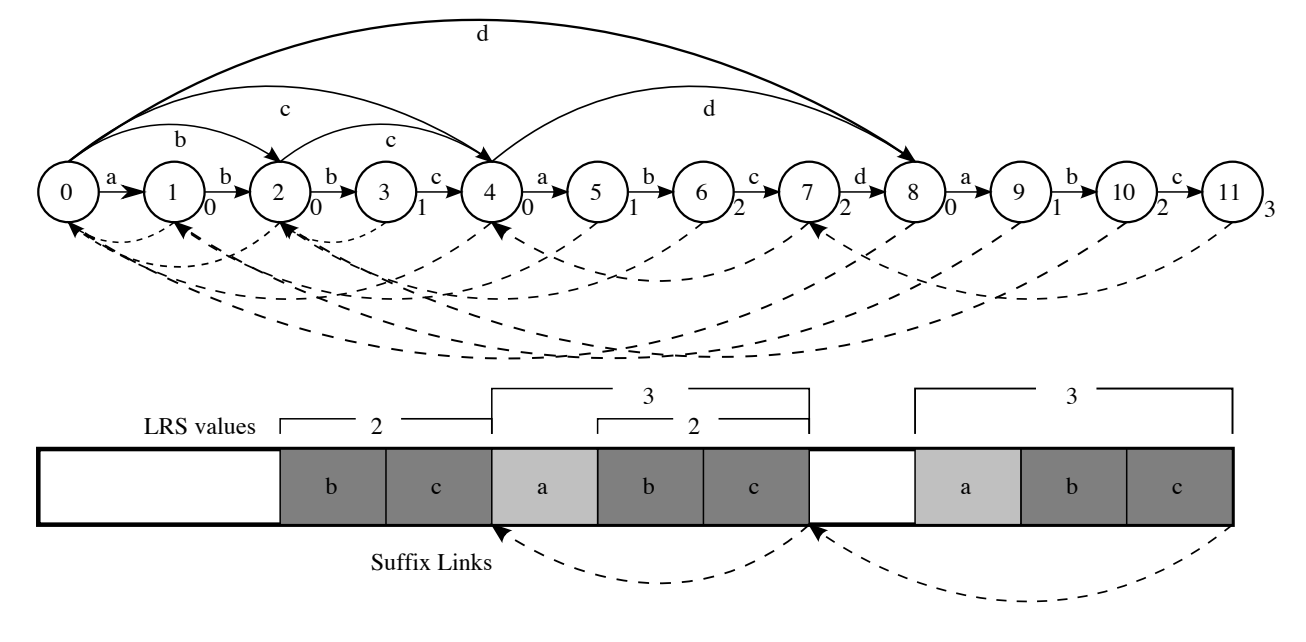}
  \caption{Reproduced from \cite{VMO}. 
	(Top) An example of the VMO structure in a symbolic signal sequence \{a, b, b, c, a, b, c, d, a, b, c\} 
  	(Bottom) A visualization of how patterns \{a,b,c\} and \{b,c\} are related to \textit{lrs} and \textit{sfx} }
  \label{fig:vmo example}
\end{figure}

\begin{figure}[h]
  \centering
  \includegraphics[width=0.75\linewidth]{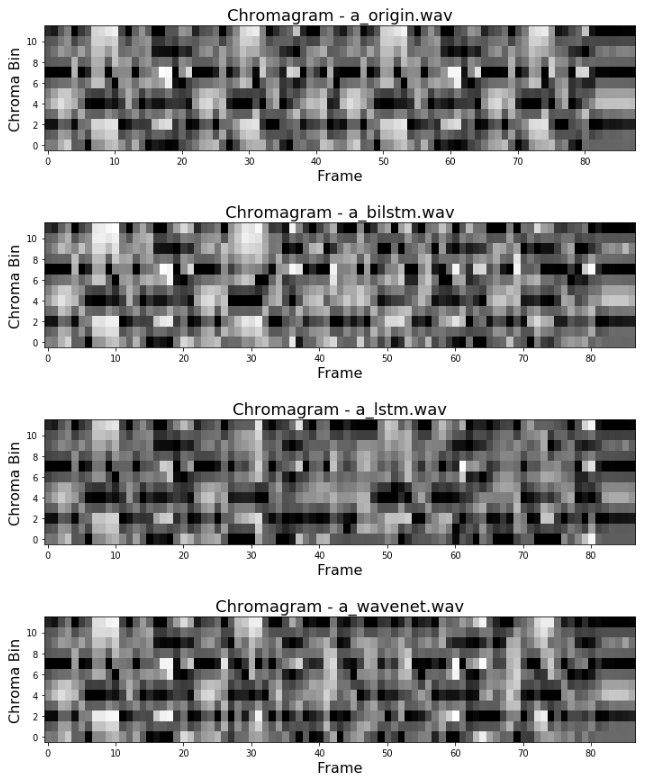}
  \caption{The chromagrams of samples with the same beginning(10 seconds)}
  \label{fig:Chromagram_example}
\end{figure}

We synthesized our generative and original midi files to waves for comparisons. We set the sample rate to 44.1 kHz and implemented \textit{short-time Fourier transform}($stft$) to get the spectrogram. Then the chromagram is obtained by folding the spectrogram into the 12 pitch classes depending on the energy. 
As shown in Fig. 8, the chromagrams of samples with the same beginning (10 seconds) are provided. From the top graph to the bottom one, each shows the original sample, bidirectional LSTM sample, unidirectional LSTM sample and the WaveNet sample.

We calculated the \textit{Information Rate} (\textit{IR})\cite{AOAna} to determine the distance threshold $\theta$. Two symbols in a time series $O$ are assigned to be the same if $|O[i]-O[j]|\leq\theta$.

Extremely high or low $\theta$ will make VMO incapable of capturing enough patterns. As shown in Fig. 9, the horizontal axis denotes $\theta$ and the vertical axis denotes \textit{IR}. A threshold is chosen (red lines) by locating the maximum \textit{IR} value. 

Finally, the patterns discovered by VMO in one samples group are shown in Fig. 10. The horizontal axis denotes the time frames and the vertical axis denotes the patterns. Graphs clearly show the disparities in different models. We will discuss these results further in Section V.
\begin{figure}[h]
  \centering
  \includegraphics[width=0.98\linewidth]{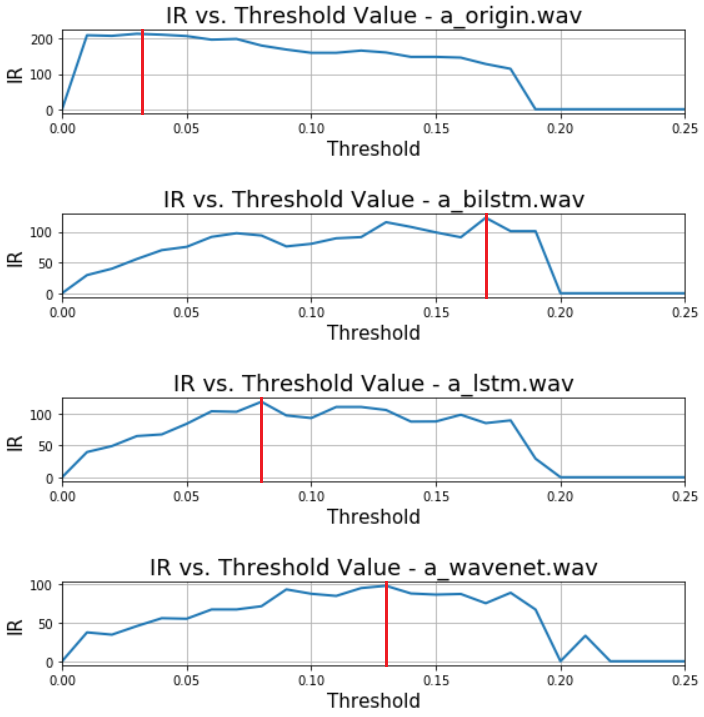}
  \caption{The \textit{IR}{-}$\theta$ graph in each sample. Red lines denote the threshold we selected in order to maximize the patterns captured in each sample.}
  \label{fig:IR_example2}
\end{figure}

\begin{figure}[h]
  \centering
  \includegraphics[width=0.9\linewidth]{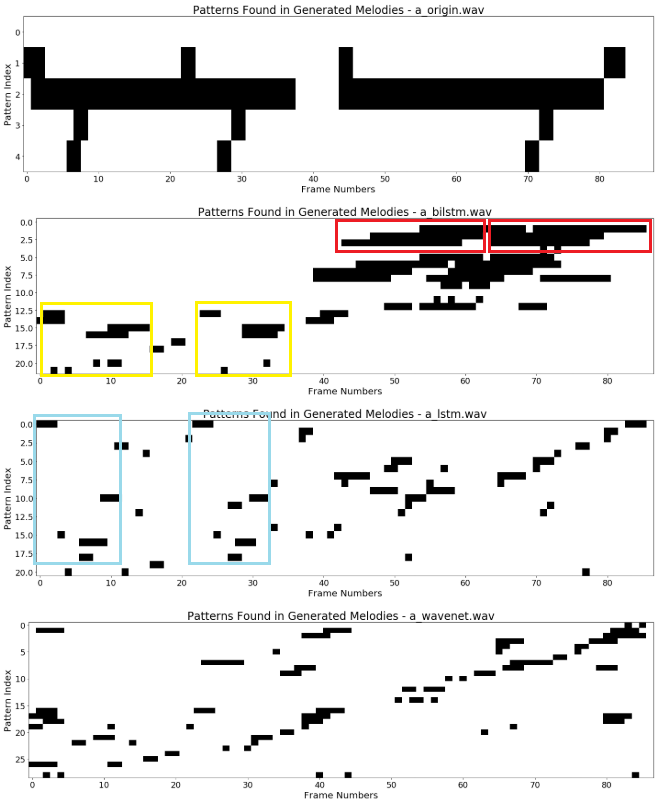}
  \caption{Patterns discovered in each sample by VMO. The horizontal lines show repetition of individual motifs. Boxes in color show repetitive patterns across motifs detected within a larger time frame. }
  \label{fig:Patterns_example}
\end{figure}

\section{Discussion}
\subsection{VMO Analysis}
VMO analysis is able to detect the motifs themselves as repeated patterns of notes, since it finds repetitions using approximately matching suffix search. The plot of these motifs over time (and sub-motifs when the lines overlap vertically) allows visual inspection of such structures. The higher level repetitions also exists in the arrangement of motifs themselves, as shown in Fig. 10 by the boxes. Since each VMO analysis optimizes the threshold of similarity for approximate suffix matching, the motifs shown in the first part of each improvisation appear slightly different. Also, the system takes into consideration also the later motif structure and adjusts its “sensitivity” so as to produce the most informative representation of the overall information in each piece.

\subsection{LSTM}
We found that music generated by LSTM model have great potential in repeating patterns. Fig. 10 shows that the unidirectional LSTM model appears a few repetitions (blue boxes), while the bidirectional LSTM model has more pattern repetitions within the time frames (red boxes and yellow boxes indicate that). Benefit from the short-term memory structure and the explicit input, it is natural for the LSTM model to capture innate structures in the dataset. 

\begin{figure}[h]
  \centering
  \includegraphics[width=0.98\linewidth]{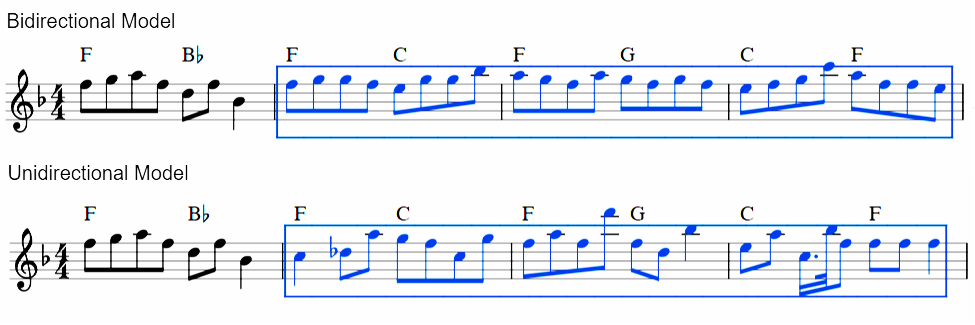}
  \caption{Comparison of results from one-way and bidirectional LSTM.}
  \label{fig:lstm}
\end{figure}

Moreover, we noticed that music generated by bidirectional LSTM is more stable and sensitive to chord changes compared to unidirectional LSTM. Fig. 11 shows an example, where the top system represents bidirectional LSTM model and the bottom system represents unidirectional LSTM model. We see that though both models can generate notes segments (of a measure) following the current chord, one-way model fails to take the chord progression as a whole. The generated note sequence tends to be more unstable, smooth, and musical for bidirectional LSTM. This is probably because the generation process includes upcoming chords as a global-structural restriction. 

The computational complexity of the LSTM model per layer is $O(nd^2)$, where $n$ is the music sequence length and $d$ is the dimension of the input. Since we implemented the chord progression condition as 25-dimension input, the extra cost of our LSTM models is reflected by the increment of dimension $d$. In practice, we did not see a problem of training efficiency.

\subsection{WaveNet}
We found that WaveNet model is able to learn some interesting rhythmic patterns. Fig. 12 shows an example, where the top staff is the (original) input sample, and the bottom staff is the generated notes. We see that both the input and output sequence contain repeated ternary rhythm patterns. Note that the grouping of triplet never breaks up. Since the step size of the generation procedure is very small, the model must have an internal long-term structural representation to capture such a rhythmic pattern.

\begin{figure}[h]
  \centering
  \includegraphics[width=0.98\linewidth]{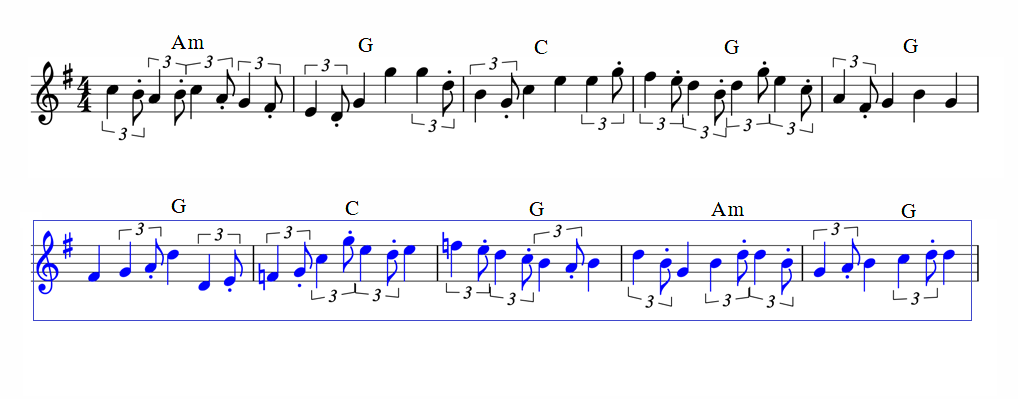}
  \caption{Generated music learns the rhythm pattern.}
  \label{fig:local_pattern}
\end{figure}

We argue that such representation comes from the explicit architectural of music structure by the stack of dilated temporal convolution layers. As shown in Fig. 4, the first layer connections can be seen as the rhythmic relationship between 1/16 beat and 1/8 beat, the second layer connections reveal the relationship between 1/8 beat and 1/4 beat, and so on so forth. In other words, the stack of dilated layers happens to agree with the hierarchical rhythmic structure of music composition. At least from the perspective of rhythm generation, WaveNet is more suitable for symbolic music generation than acoustic generation since sounds involve less hierarchical structures.

Although dilated temporal-CNN can improve the performance significantly rhythmically, it loses some structural features in the real-world music pieces. As shown in Fig. 10, it has neither the obvious repetition throughout the whole music nor the echo among music phrases. Also, unfortunately, the current WaveNet architecture is restricted to one-way music generation. It would be great to develop a bidirectional WaveNet, combining the power of explicit structure modeling and global chord conditions.

The computational complexity of dilated temporal-CNN (WaveNet) per layer is $O(knd^2)$, where $k$ is the kernel size, $n$ is the music sequence length and $d$ is the dimension of the input. Similarly to the analysis of LSTM models, the increment of dimension $d$ by the implementation of chord progression condition will not affect the training efficiency.

\section{Conclusion and Future Work}
In this paper, we compared two representative models for conditioned symbolic music generation. In the model design, we first modified WaveNet for symbolic music generation. Then, we proposed bidirectional structures in the LSTM model and further improved the performance. We conducted a subjective evaluation in our experiments, which let subjects to judge the generated samples by interactivity, complexity, and structure. We conducted \textit{Information Dynamics} analysis using \textit{Variable Markov Oracle} in order to capture the effect of different neural models on encoding of longer terms music structure. Such structure is important for human appreciation of music as explained by anticipation theories of music \cite{Sweet}. We were able to analyze the results using motif visualization technique that reveals salient repetition structures in the original versus generated output. The result shows that two critical factors largely improve the model performance: 1) a stack of dilated convolution layers which explicitly encodes the structural dependency of melody sequence, and 2) the incorporation of chord progression as a global structure constraint. In the future, we plan to combine these two factors and develop a bidirectional WaveNet for music generation.


\begin{thebibliography}{00}
\bibitem{Briot} J. Briot, G. Hadjeres, and F. Pachet, "Deep learning techniques for music  generation – A survey", CoRR, abs/1709.01620, 2017.
\bibitem{Loy} Loy and Gareth, Composing with Computers: A Survey of Some Compositional Formalisms and Music Programming Languages, MIT Press, pp. 291—396, 1989.
\bibitem{BachBot} Liang and Feynman, “BachBot: Automatic composition in the
style of Bach chorales”, University of Cambridge, 2016.
\bibitem{DeepBach} G. Hadjeres and F. Pachet, “B DeepBach: a Steerable Model for Bach chorales generation”, Proceedings of the 34th International Conference on Machine Learning, PMLR 70:1362-1371, 2017.
\bibitem{CNNBach} C. Z. Huang, T. Cooijmans, A. Roberts, A. Courville and D. Eck, “Counterpoint by Convolution”, The 18th International Society for Music Information Retrieval Conference, 2017.
\bibitem{Lattner} S. Lattner, M. Grachten and G. Widmer, “Imposing higher-level Structure in Polyphonic Music Generation using Convolutional Restricted Boltzmann Machines and Constraints”, Journal of Creative Music Systems, vol. 2, Issue 1, March 2018
\bibitem{Verma} P. Verma and J. O. Smith, “Neural Style Transfer for Audio Spectograms”, 31st Conference on Neural Information Processing Systems, Workshop for Machine Learning for Creativity and Design, 2017.
\bibitem{StructureNet} G. Medeot, S. Cherla, K. Kosta, M. McVicar, S. Abdalla, M. Selvi, E. Rex and K. Webster, “StructureNet: INDUCING STRUCTURE IN GENERATED MELODIES”, The 19th International Society for Music Information Retrieval Conference, 2018.
\bibitem{Cope} D. Cope, “Experiments in Music Intelligence (EMI)”, Proceedings of the International Computer Music Conference, 1987.
\bibitem{position} S. Dai, Z. Zhang and G. Xia, “Music Style Transfer Issues: A Position Paper”, Proceeding of International Workshop on Musical Metacreation, 2018.
\bibitem{Hild} H. Hermann, F. Johannes and M. Wolfram, “HARMONET: A Neural Net for Harmonizing Chorales in the Style of J.S.Bach”, Proceedings of the 4th International Conference on Neural Information Processing Systems, pp. 267—287, 1991. 
\bibitem{AMMS} A. Cont, S. Dubnov and G. Assayag, “Anticipatory Model of Musical Style Imitation using Collaborative and Competitive Reinforcement Learning”, Anticipatory Behavior in Adaptive Learning Systems, 2006. 
\bibitem{VMO} C. Wang, J. Hsu and S. Dubnov, “Music Pattern Discovery with Variable Markov Oracle: A Unified Approach to Symbolic and Audio Representations”, The 16th International Society for Music Information Retrieval Conference, 2015.
\bibitem{AOAna} S. Dubnov, G. Assayag and A. Cont, "Audio Oracle Analysis of Musical Information Rate", The 5th IEEE International Conference on Semantic Computing (ICSC), pp. 567–571, 2011.
\bibitem{Wavenet} A. Oord et al., “WaveNet: A Generative Model for Raw Audio”, The 9th ISCA Speech Synthesis Workshop, 2016.
\bibitem{dilated} F. Yu and V. Koltun, “Multi-Scale Context Aggregation by Dilated Convolutions”, The 4th International Conference on Learning Representations, 2016.
\bibitem{parallelwav} A. Oord et al., “Parallel WaveNet: Fast High-Fidelity Speech Synthesis”, Proceedings of the 35th International Conference on Machine Learning, PMLR 80:3918-3926, 2018.
\bibitem{fastwav} T. L. Paine et al., “Fast Wavenet Generation Algorithm”, CoRR, abs/1611.09482, 2016.
\bibitem{wavtimbre} J. Engel et al., “Neural Audio Synthesis of Musical Notes with WaveNet Autoencoders”, The 34th International Conference on Machine Learning, 2017.
\bibitem{Kobayashi} K. Kobayashi, T. Hayashi, A. Tamamori and T. Toda, “Statistical Voice Conversion with WaveNet-Based Waveform Generation”, Interspeech, 2017.
\bibitem{Tamamori} Tamamori et al., “Speaker-Dependent WaveNet Vocoder”, Interspeech, 2017.
\bibitem{Shen} J. Shen et al., “Natural TTS Synthesis by Conditioning WaveNet on Mel Spectrogram Predictions”, IEEE International Conference on Acoustics, Speech and Signal Processing, 2018.
\bibitem{Qian} K. Qian et al., “Speech Enhancement Using Bayesian Wavenet”, Interspeech, 2017.
\bibitem{manzelli2018end} M. Rachel, T. Vijay, S. Ali and K. Brian, “An end to end model for automatic music generation: Combining deep raw and symbolic audio networks”, 2018.
\bibitem{denoise} F. G. Germain, Q. Chen and V. Koltun, “Speech Denoising with Deep Feature Losses”, CoRR, abs/1806.10522, 2018.
\bibitem{Todd} Todd, “A Connectionist Approach to Algorithmic Composition”, Computer Music Journal, vol. 13 , 1989.
\bibitem{rnn} Z. C. Lipton, “A Critical Review of Recurrent Neural Networks for Sequence Learning”, CoRR, abs/1506.00019, 2015.
\bibitem{lstm} H. Sepp and S. Jurgen, “Long short-term memory”, Neural computation, MIT Press, vol. 9, pp. 1735—1780, 1997.
\bibitem{gru} K. Cho et al., “Learning Phrase Representations using RNN Encoder-Decoder for Statistical Machine Translation”, The Conference on Empirical Methods on Natural Language Processing, 2014.
\bibitem{lstmblue} D. Eck and S. Jurgen, “Learning the Long-Term Structure of the Blues”, Artificial Neural Networks — ICANN 2002, Springer Berlin Heidelberg, pp. 284—289, 2002.
\bibitem{coca} C. Andres, R. Roseli and Z. Liang, “Generation of composed musical structures through recurrent neural networks based on chaotic inspiration”, Proceedings of the International Joint Conference on Neural Networks, pp. 3220—3226, 2011.
\bibitem{Bretan} B, Mason, W. Gil and H. Larry, “A Unit Selection Methodology for Music Generation Using Deep Neural Networks”, the 8th International Conference on Computational Creativity, 2017.
\bibitem{musicvae} A. Roberts, J. Engel and D. Eck, “Hierarchical Variational Autoencoders for Music”, The 31st Conference on Neural Information Processing Systems, Workshop for Machine Learning for Creativity and Design, 2017.
\bibitem{resnet} K. He, X. Zhang, S. Ren and J. Sun, “Deep Residual Learning for Image Recognition”, IEEE Conference on Computer Vision and Pattern Recognition, 2016.
\bibitem{Nottingham} Nottingham Music Database, http://abc.sourceforge.net/NMD/, 2013.
\bibitem{anova} S. Vineeta, R. R. Kumar and S. Richa, “Analysis of repeated measurement data in the clinical trials”, Journal of Ayurveda and integrative medicine, Elsevier, vol. 4, pp. 77, 2013.
\bibitem{Sweet} D. Huron, Sweet Anticipation: Music and the Psychology of Expectation, A Bradford Book, 2008.

\end{thebibliography}
\end{document}